\documentclass[a4paper]{article}

\usepackage{amsmath}
\usepackage{fancyhdr}
\usepackage{amsfonts}
\usepackage{lscape}
\usepackage[pdftex]{graphicx}
\usepackage{amssymb}
\usepackage[utf8]{inputenc}

\title{Diffraction of  light by a nanowire}
\author{Katarzyna Krzyżanowska, Sergey Leble, leble@mifgate.pg.gda.pl\\
\\
  Gda\'nsk University of Technology, \\
ul. G. Narutowicza 11/12, 80-952 Gda\'nsk, Poland\\}
\begin{document}%

 \maketitle

\begin{abstract}
 A general scattering problem of a plane electromagnetic wave on an infinite  cylindrical rod is formulated and solved in a form of Bessel functions series expansion. The conductivity account via Ohm law directly in Maxwell equation leads to complex wavenumber and hence the complex arguments of Bessel functions inside the cylinder. The general formula for averaged by period Pointing vector is derived. For numerical calculations asymptotics of Bessel functions are used. Dependence of scattered wave intensity as function of angle and frequency is presented for different values of the rod radius. 

\textit{Keywords:} electromagnetic waves, scattering problem; conducting cylinder, nanowires;
state;  
PACS ;
\end{abstract}

\section{Introduction} 

There is a long history related to  problem of light scattering on dielectric and conducting bodies, see e.g. \cite{book}. Recent investigation  apply the theoretical results to a problems of nanowires parameters determination from optic measurements \cite{B}.
The case of dielectric rod is studied in \cite{Keam}

In this manuscript we develop the famous Mie theory to the case of infinite conducting rod of radius $r_0$ in somewhat manner similar to the cited \cite{Keam}, but widen the problem statement and its solution.  The conductivity is introduced via direct account of Ohm law in Maxwell-Ampere  equation, while the dielectric  dielectric permittivity supposed to be a real function of coordinates. We state  and solve the problem in cylindrical coordinetes.  The conductivity and  permittivity supposed to be a function of radial variable to describe eventual trasitional domain in a vicinity of the rod.  Magnetic properties are trivial: the  magnetic permeability is considered as a constant, equal to 1.

 We apply the theory to   case related to \cite{B}, using appropriate asymptotics of cylindrical functions, namely we consider scattering by thin  ($\lambda>>r_0$)   conducting rod (having in mind for example  scattering of light on semiconductor  nanowires).

\section{ Electromagnetic waves in conducting medium}

\subsection{Electromagnetic wave equation in cylindrically stratified medium}

The Maxwell equations for medium without space charges in the Lorentz-Heaviside's unit system (c - the velocity of light in vacuum)
\begin{subequations}\label{eq_Maxwell}
\begin{align}
\nabla\cdot \mathfrak{D}  = 0, \\
\nabla\times\mathfrak{E} = -\frac{1}{c}\frac{\partial\mathfrak{B}}{\partial t},\\
\nabla\cdot\mathfrak{B} = 0,\\
\nabla\times\mathfrak{H} = \frac{1}{c}\frac{\partial\mathfrak{D}}{\partial t}+\frac{4\pi}{c}\mathfrak{J},
\end{align}
\end{subequations}
where the standard set of electric $\mathfrak{D},\mathfrak{E} $ and magnetic
$\mathfrak{H}, \mathfrak{B},$  fields is used.
The material equations (of state) are assumed as for  isotropic medium. We also restrict ourselves by fixed frequency of incident and, hence, scattered wave:
\begin{subequations}\label{eq_assume}
\begin{align}
\mathfrak{D} = \epsilon \mathfrak{E} \\
\mathfrak{H} = \frac{1}{\mu}\mathfrak{B},
\end{align}
\end{subequations}
where the dielectric permittivity supposed to be a function of coordinates, while magnetic permeability - constant, 
\begin{eqnarray}
\epsilon=\epsilon(\vec{r}), \\
\mu=const.
\end{eqnarray}
The current density for this isotropic case is given by the simple version of  Ohm's law:
\begin{equation}\label{Ohm_law}
\mathfrak{J}=\sigma \mathfrak{E},
\end{equation}
where
\begin{equation}
\sigma = \sigma(\vec{r}).
\end{equation}
Indtroduce the fields  complex  amplitudes:
\begin{subequations}\label{complex_fields}
\begin{align}
\mathfrak{E}=\frac{1}{2}\left[\mathbf{E}(\vec{r})\exp{(i\omega t)}+\mathbf{E}^*(\vec{r})\exp{(-i\omega t)}\right],\\
\mathfrak{D}=\frac{1}{2}\left[\mathbf{D}(\vec{r})\exp{(i\omega t)}+\mathbf{D}^*(\vec{r})\exp{(-i\omega t)}\right],\\
\mathfrak{B}=\frac{1}{2}\left[\mathbf{B}(\vec{r})\exp{(i\omega t)}+\mathbf{B}^*(\vec{r})\exp{(-i\omega t)}\right],\\
\mathfrak{H}=\frac{1}{2}\left[\mathbf{H}(\vec{r})\exp{(i\omega t)}+\mathbf{H}^*(\vec{r})\exp{(-i\omega t)}\right].
\end{align}
\end{subequations}
Insert condtitions \eqref{complex_fields}, \eqref{Ohm_law} and \eqref{eq_assume} into the Maxwell's equations \eqref{eq_Maxwell}, then the following set of  equations for amplitudes is obtained:
\begin{subequations}\label{eq_Maxwell2}
\begin{align}
\nabla\cdot\epsilon\mathbf{E} = 0 \label{eq_Maxwell2_a}, \\
\nabla\times\mathbf{E} = -\frac{i\omega}{c}\mathbf{B}, \label{eq_Maxwell2_b}\\
\nabla\cdot\mathbf{B} = 0 \label{eq_Maxwell2_c}, \\
\nabla\times\mathbf{B} = \left(\frac{i\omega \mu \epsilon}{c}+\frac{4\pi \mu}{c} \sigma \right)\mathbf{E} \label{eq_Maxwell2_d},
\end{align}
\end{subequations}
as well as correspondent conjugated one.

Let us derive wave equation from the  set of equation \eqref{eq_Maxwell2}, differentiating as \begin{equation}\label{wav_eq}
\nabla \times \nabla \times \mathbf{E}= \left( \frac{\omega^2 \mu \epsilon}{c^2}-i \frac{4 \pi \omega \mu}{c^2} \sigma \right) \mathbf{E}.
\end{equation}
The conjugate set of equations to \eqref{eq_Maxwell2} yields  just a conjugation to  \eqref{wav_eq}.

Denote $\left( \frac{\omega^2 \mu \epsilon}{c^2}-i \frac{4 \pi \omega \mu}{c^2} \sigma \right)$ as $k^2(\vec r)$.The left side of equation \eqref{wav_eq} is equal to
\begin{equation}
\nabla \times \nabla \times \mathbf{E}= \nabla\left( \nabla \cdot \mathbf{E} \right) - \Delta \mathbf{E}. 
\end{equation}
From the equation \eqref{eq_Maxwell2_a}
\begin{equation*}
\nabla\cdot\epsilon\mathbf{E} =\epsilon \nabla \cdot \mathbf{E} + \mathbf{E} \cdot \nabla \epsilon = 0,
\end{equation*}
so
\begin{equation}
\nabla \cdot \mathbf{E} = - \frac{ \mathbf{E} \cdot \nabla \epsilon}{\epsilon} = - \mathbf{E} \cdot \frac{\nabla \epsilon}{\epsilon} ,
\end{equation}
\begin{equation}
-  \nabla \left(\mathbf{E} \cdot \frac{\nabla \epsilon}{\epsilon}\right) - \Delta \mathbf{E}=k^2(\vec{r}) \mathbf{E}.
\end{equation}

A statement of standard scattering problem in such case includes boundary conditions at infinity as prescribed asymptotics.

\subsection{Energy density flux}

The averaged flux density throughout the surface outside the rod is expressed via Pointing vector:
\begin{equation}
\mathfrak{S}_{scat}= \frac{c}{4 \pi} \frac{1}{\tau} \int_0^\tau dt \mathfrak{E}_{scat} \times \mathfrak{B}_{scat}.
\end{equation}
The flux density in terms of  the introduced  complex amplitudes \eqref{complex_fields}: 
\begin{equation*}
\mathfrak{E} \times \mathfrak{B}= 
\frac{1}{4}\left\{\mathbf{E} \times \mathbf{B} \exp{(2i\omega t)} +  \mathbf{E}^* \times \mathbf{B}^*  \exp{(-2i\omega t)} +  2 \Re{\left[\mathbf{E} \times \mathbf{B}^*\right]} \right\}.
\end{equation*}
Integrals of exponent function over the time period gives $0$. Hence:
\begin{equation}
\mathfrak{S}= \frac{c}{16 \pi} \Re{\left[\mathbf{E} \times \mathbf{B}^*\right]}.
\end{equation}

\section{Maxwell equations in cylindrical coordinates}

\subsection{Description of phenomena and its geometry}

The topic of further considerations is diffraction of the light by an infinitely long rod, hence all differential operators has a cylidrical symmetry and cylindrical variables  will be used in cylindrical coordinates $r,\varphi,z$ . In the context we restrict ourselves by the case  of $ \epsilon(\vec r)= \epsilon(r), \sigma(\vec r)=\sigma(r)$. Moreover incidental wave is a plane wave coming from infinity to $0$ along $x$ axis what causes that in general the phenomena have a symmetry with respect to $XZ$ plane, thus all derivatives over $z$  vanish $\frac{\partial }{\partial z} \equiv 0$.  It means also, that all solutions have to satisfy following condition $f(\varphi)=f(-\varphi)$.

\subsection{Cylindrical coordinates}
For the clarity all neccesary formulas, which will be used during calculations, are introduced in present section in terms of standard notations and unit vectors $  \hat{r}, \hat{\varphi}, \hat{z}$ .
Gradient in cylindrical coordinates:
\begin{equation}\label{gradient_cyl}
\nabla = \frac{\partial}{\partial r} \hat{r}+\frac{1}{r}\frac{\partial}{\partial \varphi} \hat{\varphi}+\frac{\partial}{\partial z} \hat{z},
\end{equation}
Divergence in cylindrical coordinates:
\begin{equation}\label{divergence_cyl}
\nabla \cdot \mathbf{E}= \frac{1}{r} \frac{\partial}{\partial r} \left( r E_r \right) +\frac{1}{r}\frac{\partial E_\varphi}{\partial \varphi} +\frac{\partial E_z}{\partial z}, 
\end{equation}
The curl operator:
\begin{equation}\label{rotation_cyl}
\nabla\times\mathbf{E} = \left(\frac{1}{r} \frac{\partial E_z}{\partial \varphi} - \frac{\partial E_\varphi}{\partial z}\right) \hat{r} + \left( \frac{\partial E_r}{\partial z}-\frac{\partial E_z}{\partial r} \right) \hat{\varphi} + \frac{1}{r} \left( \frac{\partial }{\partial r} \left(r E_\varphi\right)-\frac{\partial E_r}{\partial \varphi} \right) \hat{z},
\end{equation}
Laplacian (of a scalar field) in cylindrical coordinates:
\begin{equation}\label{laplacian_cyl}
\Delta = \frac{1}{r} \frac{\partial}{\partial r} (r \frac{\partial}{\partial r}) + \frac{1}{r^2} \frac{\partial^2}{\partial \varphi^2} + \frac{\partial^2}{\partial z^2}.
\end{equation}
Laplacian of a vector field $\mathbf{E}$ in cylindrical coordinates:
\begin{equation}\label{vec_laplacian_cyl}
\Delta \mathbf{E} =\left( \Delta E_r - \frac{E_r}{r^2} - \frac{2}{r^2} \frac{\partial E_\varphi}{\partial \varphi}  \right) \hat{r}+ \left( \Delta E_\varphi - \frac{E_\varphi}{r^2} + \frac{2}{r^2} \frac{\partial E_r}{\partial \varphi}  \right) \hat{\varphi}+\Delta E_z \hat{z}.
\end{equation}

\subsection{Generalized Helmnoltz equation and division of variables}

Using formluas \eqref{laplacian_cyl} and \eqref{vec_laplacian_cyl} leads to the following equations for each component of the electric field amplitude:
\begin{subequations}\label{wav_eq_r_fi_z}
\begin{align}
 - \left( \frac{1}{r} \frac{\partial}{\partial r} (r \frac{\partial E_r}{\partial r}) + \frac{1}{r^2} \frac{\partial^2 E_r}{\partial \varphi^2} - \frac{E_r}{r^2} - \frac{2}{r^2} \frac{\partial E_\varphi}{\partial \varphi} + \frac{\partial}{\partial r} \left(E_r \frac{ \partial \ln \epsilon (r)}{\partial r}\right) \right) \hat{r} = k^2(r) E_r \hat{r} \label{wav_eq_r} \\
 - \left(  \frac{1}{r} \frac{\partial}{\partial r} (r \frac{\partial E_\varphi}{\partial r}) + \frac{1}{r^2} \frac{\partial^2 E_\varphi}{\partial \varphi^2}   - \frac{E_\varphi}{r^2} + \frac{2}{r^2} \frac{\partial E_r}{\partial \varphi} + \frac{1}{r}\frac{\partial E_r}{\partial \varphi} \frac{ \partial \ln \epsilon (r)}{\partial r} \right) \hat{\varphi} = k^2(r) E_\varphi \hat{\varphi} \label{wav_eq_fi} \\
- \left(  \frac{1}{r} \frac{\partial}{\partial r} (r \frac{\partial E_z}{\partial r}) + \frac{1}{r^2} \frac{\partial^2 E_z}{\partial \varphi^2}   \right) \hat{z} =k^2(r) E_z\hat{z} \label{wav_eq_z}
\end{align}
\end{subequations}
The last equation for $\hat{z}$ coordinate and $E_z$ component \eqref{wav_eq_z} is independent of any other components, that allows to use it as basic  for further considerations. The variables are separated by substituting $E_z(r, \varphi) = R(r)\Phi(\varphi)$, hence
\begin{equation}
\frac{1}{R} r \frac{\partial R}{\partial r}+ \frac{1}{R} r^2  \frac{\partial^2 R}{\partial r^2} +k^2(r) r^2=-\frac{1}{\Phi} \frac{\partial^2 \Phi}{\partial \varphi^2} = \nu^2,
\end{equation}
where $\nu$ is a constant of separation. Consider equation for $\Phi$
\begin{equation}
\frac{\partial^2 \Phi}{\partial \varphi^2} = -\nu^2 \Phi.
\end{equation}
General solution  of it is $\Phi (\varphi) = A_1 \exp{(i\nu \varphi)}+A_2 \exp{(-i\nu \varphi)}$, however the symmetry via $XZ$ plane reduces this solution, to 
\begin{equation}
\Phi (\varphi) = A \exp{(i\nu \varphi)}. 
\end{equation}

Function $\Phi (\varphi)$ must be  continuous, and it has to satisfy the condition of periodicity $\Phi(\varphi)=\Phi(\varphi+2\pi)$. It means that the parameter $\nu$ is an integer.  Next, for each $\nu$
\begin{equation}\label{coordinate_eq_z}
r^2  \frac{\partial^2 R}{\partial r^2}+r \frac{\partial R}{\partial r} +(k^2(r) r^2- \nu^2)R(r)=0.
\end{equation}

Despite deceptive similarity to the Bessel equation of integer order, the equation above is  different because of a parameter's $k(r)$ dependence on radius coordinate. The solution of \eqref{coordinate_eq_z} involves unknown constants, so the constant $A$ can be omitted. Then the basic expansion is defined by correspondent special functions $R_\nu (r)$
\begin{equation}\label{Ez}
E_z = \sum_{\nu=-\infty}^{\infty} R_\nu (r) \exp{(i\nu \varphi)}.  
\end{equation} 

Let us expand the rest components in Fourier series in $\exp{(i\nu \varphi)}$:
\begin{equation}\label{Er}
	E_r=\sum_{\nu=-\infty}^{\infty} S_{\nu}(r) \exp{(i\nu \varphi)},
\end{equation}
\begin{equation}\label{Ephi}
	E_{\varphi}=\sum_{\nu=-\infty}^{\infty} T_{\nu}(r) \exp{(i\nu \varphi)},
\end{equation}
because of the cyclic symmetry of the   field. Plugging it into \eqref{wav_eq_r} and \eqref{wav_eq_fi} yields:
\begin{subequations}
\begin{align}
 \sum_{\nu} \left[ \frac{1}{r} \frac{\partial S_{\nu} }{\partial r} +  \frac{\partial^2 S_{\nu}}{\partial r^2} - \frac{\nu^2}{r^2}  S_{\nu} - \frac{S_{\nu}}{r^2} - \frac{i 2 \nu}{r^2} T_{\nu} + \frac{\partial S_{\nu}}{\partial r} \frac{ \partial \ln \epsilon (r)}{\partial r} +S_{\nu} \frac{ \partial^2 \ln \epsilon (r)}{\partial r^2}  + k^2(r) S_{\nu} \right] \exp{(i\nu \varphi)}=0, \\
\sum_{\nu} \left[ \frac{1}{r} \frac{\partial T_{\nu} }{\partial r} +  \frac{\partial^2 T_{\nu}}{\partial r^2} - \frac{\nu^2}{r^2}  T_{\nu} - \frac{T_{\nu}}{r^2} + \frac{i 2 \nu}{r^2} S_{\nu} + \frac{i \nu}{r}S_{\nu} \frac{ \partial \ln \epsilon (r)}{\partial r} + k^2(r) T_{\nu} \right] \exp{(i\nu \varphi)}=0.
\end{align}
\end{subequations}
Taking the linear independence of the exponents into account, and
rearranging it, one have
\begin{subequations}\label{coordinate_eq_r_fi}
\begin{align}
\frac{\partial^2 S_{\nu}}{\partial r^2}  + \left(\frac{1}{r} +  \frac{ \partial \ln \epsilon (r)}{\partial r}\right)  \frac{\partial S_{\nu} }{\partial r} +\left(\frac{ \partial^2 \ln \epsilon (r)}{\partial r^2} - \frac{\nu^2}{r^2}  - \frac{1}{r^2} +k^2(r) \right) S_{\nu} - \frac{i 2 \nu}{r^2} T_{\nu}=0 \label{coordinate_eq_r_fi_a},\\
\frac{\partial^2 T_{\nu}}{\partial r^2}+ \frac{1}{r} \frac{\partial T_{\nu} }{\partial r} +\left( - \frac{\nu^2}{r^2}   - \frac{1}{r^2}+ k^2(r)\right)T_{\nu} + \left(\frac{i 2 \nu}{r^2} + \frac{i \nu}{r} \frac{ \partial \ln \epsilon (r)}{\partial r} \right) S_{\nu} =0 \label{coordinate_eq_r_fi_b}.
\end{align}
\end{subequations}
Next, let us use the Gauss law from Maxwell equations \eqref{eq_Maxwell2_a} and divergence \eqref{divergence_cyl}  to obtain the relationship between functions $S_{\nu}(r)$ and $T_{\nu}(r)$: 
\begin{equation}
\frac{1}{r} \frac{\partial r E_r}{\partial r} + E_r \frac{\partial \ln{\epsilon(r)}}{\partial r}+\frac{1}{r} \frac{\partial E_\varphi}{\partial \varphi}=0.
\end{equation}
Combining the results, we arrive at
\begin{equation}\label{T}
T_\nu = \frac{i }{\nu} \left[ r \frac{\partial S_\nu}{\partial r}+  \left( 1+ r \frac{\partial \ln{\epsilon (r)}}{\partial r} \right) S_\nu\right]
\end{equation}
Substitute formula for $T_\nu$ into \eqref{coordinate_eq_r_fi_a}
\begin{equation}\label{S_eq}
\frac{\partial^2 S_{\nu}}{\partial r^2}  + \left(\frac{3}{r} +  \frac{ \partial \ln \epsilon (r)}{\partial r}\right)  \frac{\partial S_{\nu} }{\partial r} +\left(\frac{ \partial^2 \ln \epsilon (r)}{\partial r^2}+ \frac{2}{r} \frac{ \partial \ln \epsilon (r)}{\partial r} - \frac{\nu^2}{r^2}  + \frac{1}{r^2} +k^2(r) \right) S_{\nu}=0 
\end{equation}
Finally we have a representation for  general solution of the problem via $ S_{\nu}$. Let us pick up formulas expressing all components of the
electromagnetic $\mathbf{E}$,$\mathbf{B}$ field. Using the Maxwell equation \eqref{eq_Maxwell2_b} expresses the magnetic field by
\begin{subequations}
\begin{align}
B_z=\frac{ic}{\omega} \frac{1}{r} \left(  \frac{\partial}{\partial r} \left(  r E_\varphi \right) - \frac{\partial E_r}{\partial \varphi} \right), \\
B_r = \frac{ic}{\omega} \frac{1}{r} \frac{\partial E_z}{\partial \varphi}, \\
B_\varphi = -\frac{i c}{\omega} \frac{\partial E_z}{\partial r}.
\end{align}
\end{subequations}
After substitution of expansions, we obtain
\begin{subequations}\label{B}
\begin{align}
B_z=\frac{ic}{\omega} \sum_{\nu=-\infty}^{\infty} \left(      \frac{\partial T_\nu(r)}{\partial r} + \frac{1}{r} T_\nu(r)- \frac{i \nu}{r}S_\nu(r) \right) \exp{(i\nu\varphi)} ,\label{Bz} \\
B_r = -\frac{c}{\omega} \frac{1}{r} \sum_{\nu=-\infty}^{\infty} \nu R_\nu(r)  \exp{(i \nu \varphi )},\label{Br} \\
B_\varphi = -\frac{i c}{\omega} \sum_{\nu=-\infty}^{\infty} \frac{\partial R_\nu (r)}{\partial r} \exp{(i \nu \varphi )}\label{Bphi}.
\end{align}
\end{subequations}
It is possible to check that plugging \eqref{B} and expressions $E_r$, $E_\varphi$, $E_z$ into \eqref{eq_Maxwell2_d} with  assistance of relation $T$ of $S$ \eqref{T}, leads to the differential equations \eqref{coordinate_eq_r_fi} and \eqref{coordinate_eq_z}, which were derived from Maxwell equations set.

\section{Formulation and solution of scattering problem}

\subsection{Plane wave in cylindrical coordinates}
Let outside the semiconductive rod  vacuum. The medium is excited by a plane wave that is coming from infinity to zero along $x$ axis:
 \begin{equation}
\exp{(i k_{out} r \cos{\varphi})}=\sum_{\nu=-\infty}^{\infty} i^\nu J_\nu (k_{out} r) \exp{(i \nu \varphi)},
\end{equation}
which is identified with the incident polarised wave:
\begin{equation}
\mathbf{E}=E_z \hat{z}= \hat{z} \sum_\nu i^\nu J_\nu (k_{out} r) \exp{(i \nu \varphi)}.
\end{equation}
The correlated perpendicular magnetic field is given by
\begin{equation}
\mathbf{B}=B_y \hat{y}= \hat{y} \sum_\nu i^\nu J_\nu (k_{out} r) \exp{(i \nu \varphi)}.
\end{equation}
Find projections of $\mathbf{B}$ on the rod surface
\begin{subequations}
\begin{align}
B_r=\hat{r} \cdot \mathbf{B} = \hat{r} \cdot \hat{y} B_y= (cos{(\varphi)}\hat{x}+sin{(\varphi)}\hat{y})\cdot \hat{y} B_y = sin{(\varphi)} B_y \\
B_\varphi=\hat{\varphi} \cdot \mathbf{B} = \hat{\varphi} \cdot \hat{y} B_y= (-sin{(\varphi)}\hat{x}+cos{(\varphi)}\hat{y})\cdot \hat{y} B_y = cos{(\varphi)} B_y.
\end{align}
\end{subequations}
Use Eulers formula; because $\sum\limits_{\nu=-\infty}^{\infty}$ - summation index can be easily switched $\nu\equiv\nu+1$ or $\nu\equiv\nu-1$ \\
\begin{align*}
B_r=  \frac{\exp{(i \varphi)}-\exp{(-i \varphi)}}{2i} \sum\limits_{\nu} i^\nu J_\nu (k_{out} r) \exp{(i \nu \varphi)} = 
-\sum\limits_{\nu} \nu \frac{   i^{\nu}}{k_{out}r} J_{\nu} (k_{out} r)  \exp{(i \nu \varphi)},
\end{align*}
Similarily\\
\begin{align*}
B_\varphi=   - \sum\limits_{\nu}  \frac{  i^{\nu+1}}{k_{out}} \frac{ \partial J_{\nu} (k_{out} r)}{\partial r} \exp{(i \nu\varphi)}.
\end{align*}

\subsection{Permitivity index as a step function}
Working in polar coordinates, we divide the half axis $[0,\infty)$ to two domains $[0,a]$ - points iside a rod  and $[a,\infty)$ with constant parameters $\epsilon$ and $\sigma$.
\subsubsection{Complete solution outside rod}
Solution outside the rod consists of two fields: incidental wave denoted as $\mathfrak{E}_{inc}$ and scattered one $\mathfrak{E}_{scat}$.
\begin{equation}
\mathfrak{E}_{out} = \mathfrak{E}_{inc}+\mathfrak{E}_{scat},
\end{equation}
\begin{equation}
\mathfrak{B}_{out} = \mathfrak{B}_{inc}+\mathfrak{B}_{scat}.
\end{equation}
Incidental wave which has been already considered in the previous chapter, is well knowm polarised plane wave. Now, general form of the refracted wave will be derived by solving appropriate differential equations. We use the fillowing choice for simplicity: 
\begin{subequations}
\begin{align}
\epsilon(r) = 1,\,
\mu=1,\,
\sigma(r) = 0,
\end{align}
\end{subequations}
hence
\begin{equation}
k^2(r)= k^2_{out},
\end{equation}
all quantities do not depend on $r$ and are purely real. Taking above conditions into account the equations are simplified. Equation \eqref{coordinate_eq_z} is a Bessel equation of integer order, whose solutions are Bessel functions of the first and the second kind.
\begin{equation}
R_\nu= A_1 J_{\nu}(k_{out} r) + A_2 Y_\nu(k_{out} r).
\end{equation}
The equation \eqref{S_eq} reduces to the form:
\begin{equation}
\frac{\partial^2 S_{\nu}}{\partial r^2}  + \frac{3}{r}  \frac{\partial S_{\nu} }{\partial r} +\left( k^2_{out} - \frac{\nu^2}{r^2}  + \frac{1}{r^2}\right) S_{\nu}=0. 
\end{equation}
Substitution
$
S_\nu=\frac{1}{r} s_\nu (r)
$
leads again to the well known Bessel equation
\begin{equation}
r^2  \frac{\partial^2 s_\nu}{\partial r^2}+r \frac{\partial s_\nu}{\partial r} +(k^2_{out} r^2- \nu^2) s_\nu=0,
\end{equation}
so the solution is:
\begin{equation}
S_\nu=\frac{1}{r} \left(  A_3 J_{\nu}(k_{out} r) + A_4 Y_\nu(k_{out} r)  \right).
\end{equation}
Substituting it into \eqref{T} we get formula for:
\begin{equation}
T_\nu=\frac{i}{\nu} \left(  A_3 \frac{ \partial J_{\nu}(k_{out} r)}{\partial r} + A_4 \frac{ \partial Y_\nu(k_{out} r)}{\partial r}  \right)
\end{equation}
Both of $J_\nu$ and $Y_\nu$ are bounded for large arguments, hence general solutions of electric contributions at infinity are:
\begin{subequations}
\begin{eqnarray}
E_z= \sum_{\nu} \left(A_1 J_{\nu}(k_{out} r) + A_2 Y_\nu(k_{out} r)\right)  \exp{(i\nu \varphi)},  \\
E_r=\sum_{\nu} \frac{1}{r} \left(  A_3 J_{\nu}(k_{out} r) + A_4 Y_\nu(k_{out} r)  \right) \exp{(i\nu \varphi)}, \\
E_{\varphi}=\sum_{\nu} \frac{i}{\nu} \left(  A_3 \frac{ \partial J_{\nu}(k_{out} r)}{\partial r} + A_4 \frac{ \partial Y_\nu(k_{out} r)}{\partial r}  \right) \exp{(i\nu \varphi)}.
\end{eqnarray}
\end{subequations}
Simplifying we got all components of magnetic field:
\begin{subequations}
\begin{eqnarray}
B_z=-\frac{ic}{\omega} \sum_\nu  \left[  A_3 \left(     \frac{i k^2_{out}}{\nu} J_\nu (k_{out} r)      \right) \right. +  \left. A_4 \left( \frac{i k^2_{out}}{\nu} Y_\nu (k_{out} r)  \right) \right] \exp{(i\nu\varphi)}, \\
B_r = -\frac{c}{\omega} \frac{1}{r} \sum_\nu \nu \left( A_1 J_{\nu}(k_{out} r) + A_2 Y_\nu(k_{out} r) \right)  \exp{(i \nu \varphi )}, \\
B_\varphi = -\frac{i c}{\omega} \sum_\nu \left(A_1 \frac{\partial J_{\nu}(k_{out} r)}{\partial r} + A_2 \frac{\partial Y_{\nu}(k_{out} r)}{\partial r}\right) \exp{(i \nu \varphi )}.
\end{eqnarray}
\end{subequations}
Now it is possible to write down explicit forms of polarized electric and magnetic field outside the cylinder.
\subsubsection{Solution inside}
Analogously to the previous section we put
\begin{subequations}
\begin{align}
\epsilon(r) = \epsilon_{in},\,
\mu=1,\,
\sigma(r) =  \sigma_{in}.
\end{align}
\end{subequations}
\begin{equation}
k^2 
= \left( \frac{\omega^2 \epsilon_{in}}{c^2}-i \frac{4 \pi \omega}{c^2} \sigma_{in} \right) = k^2_{in}
\end{equation}
In the case of stepfunctions, $k_{in}$ does not depend on radial coordinate but this time it is a complex number. It does not change the form of the solutions, the only difference is that the argument of the solutions is complex. Upper bounds of solutions have to be limited, so Neumann function and its derivatives contributions that blows up at zero must be suppressed. It gives
\begin{subequations}
\begin{eqnarray}
E_z= \sum_{\nu} A_5 J_{\nu}(k_{in} r)  \exp{(i\nu \varphi)},  \\
E_r=\sum_{\nu} \frac{1}{r}  A_6 J_{\nu}(k_{in} r) \exp{(i\nu \varphi)}, \\
E_{\varphi}=\sum_{\nu} \frac{i}{\nu}  A_6 \frac{ \partial J_{\nu}(k_{in} r)}{\partial r}  \exp{(i\nu \varphi)}.
\end{eqnarray}
\end{subequations}
\begin{subequations}
\begin{eqnarray}
B_z=-\frac{ic}{\omega} \sum_\nu  A_6    \frac{i k^2_{in}}{\nu} J_\nu (k_{in} r)           \exp{(i\nu\varphi)}, \\
B_r = -\frac{c}{\omega} \frac{1}{r} \sum_\nu \nu  A_5 J_{\nu}(k_{in} r)   \exp{(i \nu \varphi )} ,\\
B_\varphi = -\frac{i c}{\omega} \sum_\nu A_5 \frac{\partial J_{\nu}(k_{in} r)}{\partial r} \exp{(i \nu \varphi )}.
\end{eqnarray}
\end{subequations}

\subsubsection{Polarization towards $z$}

Explicit forms of polarised electric and magnetic field outside the cylinder yields:
\begin{equation}
\begin{split}
2 \mathfrak{E}_{out}= \left[\hat{z}  \sum_\nu i^\nu J_\nu (k_{out} r) \exp{(i \nu \varphi)} + \hat{z}  \sum_{\nu} \left(A_1 J_{\nu}(k_{out} r) + A_2 Y_\nu(k_{out} r)\right)  \exp{(i\nu \varphi)} \right]\exp{(i\omega t)}+
\\
+ \left[\sum_\nu (-i)^\nu J_\nu (k_{out} r) \exp{( - i \nu \varphi)} + \hat{z}  \sum_{\nu} \left(A^*_1 J_{\nu}(k_{out} r) + A^*_2 Y_\nu(k_{out} r)\right)  \exp{( - i\nu \varphi)} \right]\exp{(-i\omega t)}
\end{split}
\end{equation}
\begin{equation}
\begin{split}
2 \mathfrak{B}_{out}= \left[   -\hat{r} \sum\limits_{\nu}   i^{\nu} \nu \frac{1}{r} \frac{1}{k_{out}} J_{\nu} (k_{out} r)  \exp{(i \nu \varphi)} \right. + \\
-\hat{\varphi} \sum\limits_{\nu}  i i^{\nu} \frac{1}{k_{out}} \frac{ \partial J_{\nu} (k_{out} r)}{\partial r} \exp{(i \nu\varphi)} + \\
- \hat{r} \sum_\nu \frac{c}{\omega} \frac{1}{r} \nu \left( A_1 J_{\nu}(k_{out} r) + A_2 Y_\nu(k_{out} r) \right)  \exp{(i \nu \varphi )} + \\
- \left. \hat{\varphi} \sum_\nu  \frac{i c}{\omega} \left(A_1 \frac{\partial J_{\nu}(k_{out} r)}{\partial r} + A_2 \frac{\partial Y_{\nu}(k_{out} r)}{\partial r}\right) \exp{(i \nu \varphi )}    \right] \exp{(i\omega t)} +
\\
+\left[ -  \hat{r} \sum\limits_{\nu}   (-i)^{\nu} \nu \frac{1}{r} \frac{1}{k_{out}} J_{\nu} (k_{out} r)  \exp{( - i \nu \varphi)} \right. + \\
+\hat{\varphi} \sum\limits_{\nu}  i (-i)^{\nu} \frac{1}{k_{out}} \frac{ \partial J_{\nu} (k_{out} r)}{\partial r} \exp{( - i \nu\varphi)} + \\
- \hat{r} \sum_\nu \frac{c}{\omega} \frac{1}{r} \nu \left( A^*_1 J_{\nu}(k_{out} r) + A^*_2 Y_\nu(k_{out} r) \right)  \exp{(-i \nu \varphi )} + \\
+ \left. \hat{\varphi} \sum_\nu  \frac{i c}{\omega} \left(A^*_1 \frac{\partial J_{\nu}(k_{out} r)}{\partial r} + A^*_2 \frac{\partial Y_{\nu}(k_{out} r)}{\partial r}\right) \exp{(-i \nu \varphi )}    \right] \exp{(-i\omega t)} 
\end{split}
\end{equation}
\begin{equation}
2 \mathfrak{E}_{in} = \left[\hat{z}  \sum_{\nu} A_5 J_{\nu}(k_{in} r)  \exp{(i\nu \varphi)} \right] \exp{(i \omega t)}+
\left[\hat{z}  \sum_{\nu} A^*_5 J_{\nu}(k^*_{in} r)  \exp{(-i\nu \varphi)} \right] \exp{(-i \omega t)}
\end{equation}
\begin{equation}
\begin{split}
2 \mathfrak{B}_{in} = \left[- \hat{r}  \sum_\nu \frac{c}{\omega} \frac{1}{r} \nu  A_5 J_{\nu}(k_{in} r)   \exp{(i \nu \varphi )} - \hat{\varphi} \sum_\nu  \frac{i c}{\omega} A_5 \frac{\partial J_{\nu}(k_{in} r)}{\partial r} \exp{(i \nu \varphi )}\right]\exp{(i \omega t)}+
\\
+ \left[- \hat{r}  \sum_\nu \frac{c}{\omega} \frac{1}{r} \nu  A^*_5 J_{\nu}(k^*_{in} r)   \exp{(-i \nu \varphi )} + \hat{\varphi} \sum_\nu  \frac{i c}{\omega} A^*_5 \frac{\partial J_{\nu}(k^*_{in} r)}{\partial r} \exp{(-i \nu \varphi )}\right]\exp{(-i \omega t)}
\end{split}
\end{equation}

\subsection{Boundary conditions on a rod surface for polarizations along  $z$, }

Tangential components of electric and perpendicular contribution of magnetic field must be continious on a surface of the rod. Components toward $z$ and $\varphi$ axes are always tangential to the cylinder surface, what is universal, whatever light polarisation is. Assume the wave polarisation is directed along $z$, then:
\begin{subequations}
\begin{align}
\left.\left(\mathfrak{E}_{out}\right)_{\hat{z}}\right|_{r_0} -  \left.\left(\mathfrak{E}_{in}\right)_{\hat{z}}\right|_{r_0} = 0,  
\\
\left.\left(\mathfrak{B}_{out}\right)_{\hat{r}}\right|_{r_0} -  \left.\left(\mathfrak{B}_{in}\right)_{\hat{r}}\right|_{r_0} = 0,  
\\
\left.\left(\mathfrak{B}_{out}\right)_{\hat{\varphi}}\right|_{r_0}- \left.\left(\mathfrak{B}_{in}\right)_{\hat{\varphi}}\right|_{r_0}= 0.
\end{align}
\end{subequations}
Because $\exp{(i\omega t)}$ and $\exp{(-i \omega t)}$ are linearly independent, above conditions can be separated:
\begin{subequations}
\begin{align}
\left.\left(\mathbf{E}_{out}\right)_{\hat{z}}\right|_{r_0} - \left.\left(\mathbf{E}_{in}\right)_{\hat{z}}\right|_{r_0} = 0,  
\\
\left.\left(\mathbf{E}^*_{out}\right)_{\hat{z}}\right|_{r_0} -\left.\left(\mathbf{E}^*_{in}\right)_{\hat{z}}\right|_{r_0} = 0,  
\\
\left.\left(\mathbf{B}_{out}\right)_{\hat{r}}\right|_{r_0} - \left.\left(\mathbf{B}_{in}\right)_{\hat{r}}\right|_{r_0} = 0,  
\\
\left.\left(\mathbf{B}^*_{out}\right)_{\hat{r}}\right|_{r_0} - \left.\left(\mathbf{B}^*_{in}\right)_{\hat{r}}\right|_{r_0} = 0,  
\\
\left.\left(\mathbf{B}_{out}\right)_{\hat{\varphi}}\right|_{r_0}- \left.\left(\mathbf{B}_{in}\right)_{\hat{\varphi}}\right|_{r_0}= 0,
\\
\left.\left(\mathbf{B}^*_{out}\right)_{\hat{\varphi}}\right|_{r_0}-\left.\left(\mathbf{B}^*_{in}\right)_{\hat{\varphi}}\right|_{r_0}= 0.
\end{align}
\end{subequations}
Explicit form after substitution of the field components leads to the set of two equations, which contain three unknown constants.
\begin{subequations}
\begin{align}
A_5 J_{\nu}(k_{in} r_0)  =  ( i^\nu+  A_1) J_\nu (k_{out} r_0)    + A_2 Y_\nu(k_{out} r_0),
\\
 A_5 \left.\frac{\partial J_{\nu}(k_{in} r)}{\partial r}\right|_{r_0} = ( i^{\nu}+   A_1 ) \left.\frac{\partial J_{\nu}(k_{out} r)}{\partial r}\right|_{r_0}   +  A_2 \left.\frac{\partial Y_{\nu}(k_{out} r)}{\partial r}\right|_{r_0}.
\end{align}
\end{subequations}
It gives us formula for amplitude inside rod
\begin{equation}
A_5=\frac{i^\nu J_{\nu}(k_{out} r_0) + A_1 J_{\nu}(k_{out} r_0)+ A_2 Y_{\nu}(k_{out} r_0)}{J_{\nu}(k_{in} r_0)},
\end{equation}
and the relation between constants $A_1$ and $A_2$.
\begin{equation}\label{eq:A1_A2_rel}
A_2=-(A_1 + i^\nu) \left. \frac{  \frac{\partial}{\partial r} \left( \frac{J_\nu (k_{out} r)}{J_{\nu}(k_{in} r)}  \right)}{ \frac{\partial}{\partial r} \left( \frac{Y_\nu (k_{out} r)}{J_{\nu}(k_{in} r)}  \right)} \right|_{r_0}.
\end{equation}
As it is shown in the formula \eqref{eq:A1_A2_rel} there is a freedom in choosing explicit relation of $A_1$ and $A_2$. By direct physical intuition,  we choose constants such that 
\begin{equation}
A_2=-iA_1.
\end{equation}
Hence refracted wave would be described by a Hankel function of the first kind $H^{(+)}_\nu (k_{out} r)=J_\nu (k_{out} r)+i Y_\nu (k_{out} r)$. Hankel function tends to be a cylindrical wave at infinity, so it is a good choice. Finally,
\begin{equation}
A_5=\frac{i^\nu J_{\nu}(k_{out} r_0) + A_1 H^{(2)}_{\nu}(k_{out} r_0)}{J_{\nu}(k_{in} r_0)},
\end{equation}
\begin{equation}
A_1=- i^\nu \frac{ k_{out} \left( J_{\nu-1}(k_{out} r_0 ) - J_{\nu+1}(k_{out} r_0 ) \right) J_\nu(k_{in} r_0) - k_{in} J_\nu(k_{out} r_0) \left(J_{\nu-1}(k_{in} r_0 ) - J_{\nu+1}(k_{in} r_0 ) \right) }{ k_{out} \left( H^{(2)}_{\nu-1}(k_{out} r_0 ) - H^{(2)}_{\nu+1}(k_{out} r_0 ) \right) J_\nu(k_{in} r_0) - k_{in} H^{(2)}_\nu(k_{out} r_0) \left(J_{\nu-1}(k_{in} r ) - J_{\nu+1}(k_{in} r_0 ) \right) } . 
\end{equation}

\subsection{Properties and approximations of sum over $A_1(\nu)$}

Assume $r_0<100 \AA$. Taking into account the physical values of introduced parameters, it is possible to approximate estimated amplitude $A_1$. While the argument is small enough ($k_{out} r_0 = \frac{2 \pi r_0}{\lambda} \cong \frac{6 \cdot 100 \AA }{6000 \AA} = 0.1$), Bessel functions tend to be a monomials.
\begin{subequations}\label{J_H_approx}
\begin{align}
J_\nu(z)\approx \frac{\left(\frac{1}{2} z\right)^\nu}{\Gamma (\nu+1)}, & \mbox{        } \nu \neq -1, -2, \ldots \\
H_\nu(z)\approx - i \frac{1}{\pi} \frac{\Gamma (\nu)} {\left(\frac{1}{2} z\right)^\nu},  & \mbox{        } \Re{(\nu)}>0.
\end{align}
\end{subequations}
Next,
\begin{equation}
\left\|k_{in} r_0\right\| = \frac{r_0 \omega}{c} \sqrt[4]{\epsilon^2+ 16 \pi^2 \left(\frac{\sigma}{\omega}\right)^2}.
\end{equation}
There is one more restrictions imposed by formula \eqref{J_H_approx}, that is $\nu$ must be nonnegative. Remember that the solution $E_z$ requires sum over all integer values of $\nu$, from minus infinity to plus infinity. However, it can be easily check, that $A_1(\nu)$ is even or odd function for even or odd $\nu$ respectively:
\begin{equation}
A_1(\nu)=(-)^\nu A_1(-\nu),
\end{equation}
so the constant can be approximated as follows:
\begin{equation}
A_1 \approx -i^\nu \frac{i \pi \left(\frac{1}{2} k_{out} r_0\right)^{2\nu +2} \frac{ (\nu-1) }{\Gamma^2(\nu+1)} \left( \left(\frac{k_{in}}{k_{out}}\right)^2-1\right)}{\left(\frac{1}{2} k_{out} r_0\right)^2 - 2(\nu^2-1)}
\end{equation}
Formula above should be used very carefuly, because it is no more even or odd via $\nu$.

\subsection{Energy density flux evaluation}

The Pointing vector 
\begin{equation}
\mathfrak{S} = \frac{c}{16 \pi} \Re{\left[\mathbf{E} \times \mathbf{B}^*\right]} = \frac{c}{16 \pi} \left(  -2 \Re{\left[E_z B^*_\varphi\right]}\hat{r} +2 \Re{\left[E_z B^*_r\right]} \hat{\varphi} \right),
\end{equation}
defines the instant energy density flux. We are interested in contribution toward radial direction $\hat{r}$ for big $z$.
\begin{equation}
\left( \mathfrak{S}_{scat} \right)_{\hat{r}} = - \frac{c}{8 \pi} \Re{\left[E_{scat,z} B^*_{scat, \varphi}  \right]},
\end{equation}
Asymptotics at large $ z$ is
\begin{equation}
H_\nu^{(2)}(z) \approx \sqrt{\frac{2}{\pi z}} \exp{\left( -i z+ i \frac{\nu \pi}{2}+i\frac{\pi}{4} \right)},
\end{equation}
hence
\begin{equation*}
\begin{array}{c}
E_{scat,z}=\sum_\nu A_1(\nu)  H_\nu^{(2)}(k_{out} r)  \exp{(i \nu \varphi)}\approx\\ \sqrt{\frac{2}{\pi k_{out} r}} \exp{\left( -i k_{out} r +i\frac{\pi}{4} \right)}\sum_\nu A_1(\nu)  \exp{\left( i \frac{\nu \pi}{2}\right)}  \exp{(i \nu \varphi)},
\end{array}
\end{equation*}
\begin{equation*}
\begin{split}
B_{scat,\varphi}=-\sum_\nu \frac{i c}{\omega} A_1(\nu) \frac{\partial H_\nu^{(2)}(k_{out} r) }{\partial r}   \exp{(i \nu \varphi)} = \\
= -\sum_\nu \frac{i c}{\omega} A_1(\nu) k_{out} \frac{1}{2} \left[ H_{\nu-1}^{(2)}(k_{out} r)  - H_{\nu+1}^{(2)}(k_{out} r)   \right]  \exp{(i \nu \varphi)} \approx \\
\approx - \sqrt{\frac{2}{\pi k_{out} r}} \exp{\left( -i k_{out} r +i\frac{\pi}{4} \right)} \sum_\nu i A_1(\nu) \exp{\left( i \frac{\nu \pi}{2}\right)} \frac{1}{2} \left[ \exp{\left( - i \frac{\pi}{2}\right)} - \exp{\left( i \frac{\pi}{2}\right)}  \right]  \exp{(i \nu \varphi)}=\\
= - \sqrt{\frac{2}{\pi k_{out} r}} \exp{\left( -i k_{out} r +i\frac{\pi}{4} \right)} \sum_\nu  A_1(\nu) \exp{\left( i \frac{\nu \pi}{2}\right)} \exp{(i \nu \varphi)}
\end{split}
\end{equation*}
Denote $A_1=-i^{\nu}a_1$, then
\begin{equation*}
\sum_\nu A_1(\nu)  \exp{\left( i \frac{\nu \pi}{2}\right)}  \exp{(i \nu \varphi)}=- \sum_\nu a_1(\nu)  \exp{\left( i \nu \pi \right)}  \exp{(i \nu \varphi)},
\end{equation*}
\begin{equation}
E_{scat,z} B^*_{scat,\varphi}=-\frac{2}{\pi k_{out} r} \left\| \sum_\nu a_1(\nu)  \exp{\left(i \nu \pi\right)}  \exp{(i \nu \varphi)} \right\|^2.
\end{equation}
Therefore
\begin{equation}
\left( \mathfrak{S}_{scat} \right)_{\hat{r}} =\frac{c}{4 \pi^2} \frac{1}{k_{out} r} \sum_{\nu, \nu'} (-1)^{(\nu-\nu')}  a_1(\nu) a_1^*(\nu') \exp{(i (\nu-\nu') \varphi)}.
\end{equation}

\subsection{Polarization toward $y$, the list of formulas }
\subsubsection{Fields}
\begin{equation}
\begin{split}
2 \mathfrak{E}_{out}= \left[   -\hat{r} \sum\limits_{\nu}   i^{\nu} \nu \frac{1}{r} \frac{1}{k_{out}} J_{\nu} (k_{out} r)  \exp{(i \nu \varphi)} \right. + \\
-\hat{\varphi} \sum\limits_{\nu}  i i^{\nu} \frac{1}{k_{out}} \frac{ \partial J_{\nu} (k_{out} r)}{\partial r} \exp{(i \nu\varphi)} + \\
+ \hat{r} \sum_{\nu} \frac{1}{r} \left(  A_3 J_{\nu}(k_{out} r) + A_4 Y_\nu(k_{out} r)  \right) \exp{(i\nu \varphi)} + \\
+ \left. \hat{\varphi} \sum_{\nu} \frac{i}{\nu} \left(  A_3 \frac{ \partial J_{\nu}(k_{out} r)}{\partial r} + A_4 \frac{ \partial Y_\nu(k_{out} r)}{\partial r}  \right) \exp{(i\nu \varphi)}    \right] \exp{(i\omega t)} +
c.c 
\end{split}
\end{equation}
\begin{equation}
\begin{split}
2 \mathfrak{B}_{out}= \left[ - \hat{z}  \sum_\nu i^\nu J_\nu (k_{out} r) \exp{(i \nu \varphi)} + \hat{z}  \sum_\nu \frac{c}{\omega} \frac{ k^2_{out}}{\nu} \left[  A_3      J_\nu (k_{out} r)    +   A_4  Y_\nu (k_{out} r)   \right] \exp{(i\nu\varphi)} \right]\exp{(i\omega t)}+
c.c 
\end{split}
\end{equation}
\begin{equation}
\begin{split}
2 \mathfrak{E}_{in} = \left[ \hat{r}  \sum_{\nu} \frac{1}{r}  A_6 J_{\nu}(k_{in} r) \exp{(i\nu \varphi)} + \hat{\varphi} \sum_{\nu} \frac{i}{\nu}  A_6 \frac{ \partial J_{\nu}(k_{in} r)}{\partial r}  \exp{(i\nu \varphi)} \right]\exp{(i \omega t)}+
c.c 
\end{split}
\end{equation}
\begin{equation}
2 \mathfrak{B}_{in} = \left[\hat{z}  \sum_\nu \frac{c}{\omega} A_6    \frac{ k^2_{in}}{\nu} J_\nu (k_{in} r)    \exp{(i\nu\varphi)} \right] \exp{(i \omega t)}+
c.c 
\end{equation}

\subsubsection{Boundary conditions on a rod surface for the polarization}

\begin{subequations}
\begin{align}
\left.\left(\mathbf{B}_{out}\right)_{\hat{z}}\right|_{r_0} -  \left.\left(\mathbf{B}_{in}\right)_{\hat{z}}\right|_{r_0} = 0,\\
\epsilon_{out} \left.\left(\mathbf{E}_{out}\right)_{\hat{r}}\right|_{r_0} -  \epsilon_{in} \left.\left(\mathbf{E}_{in}\right)_{\hat{r}}\right|_{r_0} = 0,  
\\
\left.\left(\mathbf{E}_{out}\right)_{\hat{\varphi}}\right|_{r_0}-\left.\left(\mathbf{E}_{in}\right)_{\hat{\varphi}}\right|_{r_0}= 0.
\end{align}
\end{subequations}
\begin{equation}
\begin{array}{c}
\frac{ k^2_{in}}{ k^2_{out} } A_6 J_\nu (k_{in} r_0)  =  ( A_3- i^\nu \frac{ \nu}{k_{out}}) J_\nu (k_{out} r_0)     +   A_4  Y_\nu (k_{out} r_0), \\
 \frac{\epsilon_{in}}{\epsilon_{out}}   A_6 J_{\nu}(k_{in} r_0)  = (  A_3 -   \frac{ i^{\nu}\nu}{k_{out}} )J_{\nu} (k_{out} r_0)    + A_4 Y_\nu(k_{out} r_0) ,
\\
A_6  \left. \frac{ \partial J_{\nu}(k_{in} r)}{\partial r} \right|_{r0} = ( A_3 -  \frac{ i^{\nu} \nu}{k_{out}}) \left. \frac{ \partial J_{\nu} (k_{out} r)}{\partial r} \right|_{r0}   + A_4 \left. \frac{ \partial Y_\nu(k_{out} r)}{\partial r} \right|_{r0} .
\end{array}
\end{equation}
similar to the previous case,
\begin{equation*}
A_3=-iA_4.
\end{equation*}
\begin{equation}
A_3= \frac{i^\nu \nu}{k_{out}} \left. \frac   { \left(\frac{k^2_{in}}{k^2_{out}}\right) \left(\frac{\partial}{\partial r} J_\nu(k_{out} r )\right) J_\nu(k_{in} r) - J_\nu(k_{out} r) \left(\frac{\partial}{\partial r} J_\nu(k_{in} r )\right) }  { \left(\frac{k^2_{in}}{k^2_{out}}\right) \left(\frac{\partial}{\partial r} H^{(2)}_\nu(k_{out} r )\right) J_\nu(k_{in} r) - H^{(2)}_\nu(k_{out} r) \left(\frac{\partial}{\partial r} J_\nu(k_{in} r )\right)}\right|_{r_0}.
\end{equation}

\subsubsection{Energy density flux}

\begin{equation}
\mathfrak{S} = \frac{c}{16 \pi} \Re{\left[\mathbf{E} \times \mathbf{B}^*\right]} = \frac{c}{16 \pi} \left(  \frac{1}{r}2 \Re{\left[E_\varphi B^*_z\right]}\hat{r} -2 \Re{\left[E_r B^*_z\right]} \hat{\varphi} \right).
\end{equation}
We are interested again in contribution towards radial direction $\hat{r}$.
\begin{equation}
\left( \mathfrak{S}_{scat} \right)_{\hat{r}} =  \frac{c}{8 \pi r} \Re{\left[E_{scat,\varphi} B^*_{scat, z}  \right]},
\end{equation}
\begin{equation}
H_\nu^{(2)}(z) \approx \sqrt{\frac{2}{\pi z}} \exp{\left( -i z+ i \frac{\nu \pi}{2}+i\frac{\pi}{4} \right)},
\end{equation}
\begin{equation*}
\begin{split}
E_{scat,\varphi}=\sum_\nu \frac{i}{\nu} A_3(\nu) \frac{\partial H_\nu^{(2)}(k_{out} r) }{\partial r}   \exp{(i \nu \varphi)} = \\
= \sum_\nu \frac{i}{\nu} k_{out} A_3(\nu)  \frac{1}{2} \left[ H_{\nu-1}^{(2)}(k_{out} r)  - H_{\nu+1}^{(2)}(k_{out} r)   \right]  \exp{(i \nu \varphi)} \approx \\
\approx  \sqrt{\frac{2}{\pi k_{out} r}} \exp{\left( -i k_{out} r +i\frac{\pi}{4} \right)} \sum_\nu \frac{i}{\nu} k_{out}  A_3(\nu) \exp{\left( i \frac{\nu \pi}{2}\right)} \frac{1}{2} \left[ \exp{\left( - i \frac{\pi}{2}\right)} - \exp{\left( i \frac{\pi}{2}\right)}  \right]  \exp{(i \nu \varphi)}=\\
=  \sqrt{\frac{2}{\pi k_{out} r}} \exp{\left( -i k_{out} r +i\frac{\pi}{4} \right)} \sum_\nu \frac{1}{\nu} k_{out}  A_3(\nu) \exp{\left( i \frac{\nu \pi}{2}\right)} \exp{(i \nu \varphi)}
\end{split}
\end{equation*}
\begin{equation*}
B_{scat,z}= - \frac{i c}{\omega}  \sum_\nu A_3(\nu) \frac{i k^2_{out}}{\nu}  H_\nu^{(2)}(k_{out} r)  \exp{(i \nu \varphi)}\approx  \sqrt{\frac{2}{\pi k_{out} r}} \exp{\left( -i k_{out} r +i\frac{\pi}{4} \right)}\sum_\nu \frac{k_{out}}{\nu} A_3(\nu)  \exp{\left( i \frac{\nu \pi}{2}\right)}  \exp{(i \nu \varphi)},
\end{equation*}
\begin{equation}
A_3=i^\nu \nu \frac{1}{k_{out}} a_3(\nu),
\end{equation}
\begin{equation*}
\begin{split}
\sum_\nu \frac{1}{\nu} k_{out}  A_3(\nu) \exp{\left( i \frac{\nu \pi}{2}\right)} \exp{(i \nu \varphi)}=\\
=\sum_\nu \frac{1}{\nu} k_{out}   i^\nu \nu \frac{1}{k_{out}} a_3(\nu) \exp{\left( i \frac{\nu \pi}{2}\right)} \exp{(i \nu \varphi)}=\\
= \sum_\nu    a_3(\nu) \exp{\left( i \nu \pi \right)} \exp{(i \nu \varphi)},
\end{split}
\end{equation*}
\begin{equation}
E_{scat,\varphi} B^*_{scat,z}=\frac{2}{\pi k_{out} r} \left\| \sum_\nu a_3(\nu)  \exp{\left(i \nu \pi\right)}  \exp{(i \nu \varphi)} \right\|^2,
\end{equation}
\begin{equation}
\left( \mathfrak{S}_{scat} \right)_{\hat{r}} = \frac{c}{4 \pi^2} \frac{1}{k_{out} r} \left\| \sum_\nu a_3(\nu)  \exp{\left( i\nu \pi\right)}  \exp{(i \nu \varphi)} \right\|^2.
\end{equation}

\section{Results discussion}
The derived formulas for the z-polarization coincide with ones from \cite{B} if one take into account the relation \begin{equation}
\frac{k_{in}}{k_{out}}=\tilde{n},
\end{equation}
and change notations as
\begin{subequations}
\begin{align}
a_3(\nu)=a_\nu \\
a_1(\nu)=b_\nu
\end{align}
\end{subequations}
\begin{equation}
Q_{TM}= \frac{1}{k_{out} r} \left\| \sum_\nu b_\nu  \exp{\left( i\nu \pi\right)}  \exp{(i \nu \varphi)} \right\|^2=\frac{1}{k_{out} r} \sum_{\nu, \nu'} (-1)^{(\nu-\nu')}  b_\nu b^*_{\nu'} \exp{(i (\nu-\nu') \varphi)}
\end{equation}
The main target of this paper is to derive and investigate dependence of scattered wave on the material parameters and dimension of the rod. For thie purpose a code and program were elaborated.  Below we illustrate the results of calcullations by  plots (Figure 1). 
\begin{figure}
	\centering
		\includegraphics[scale=0.25]{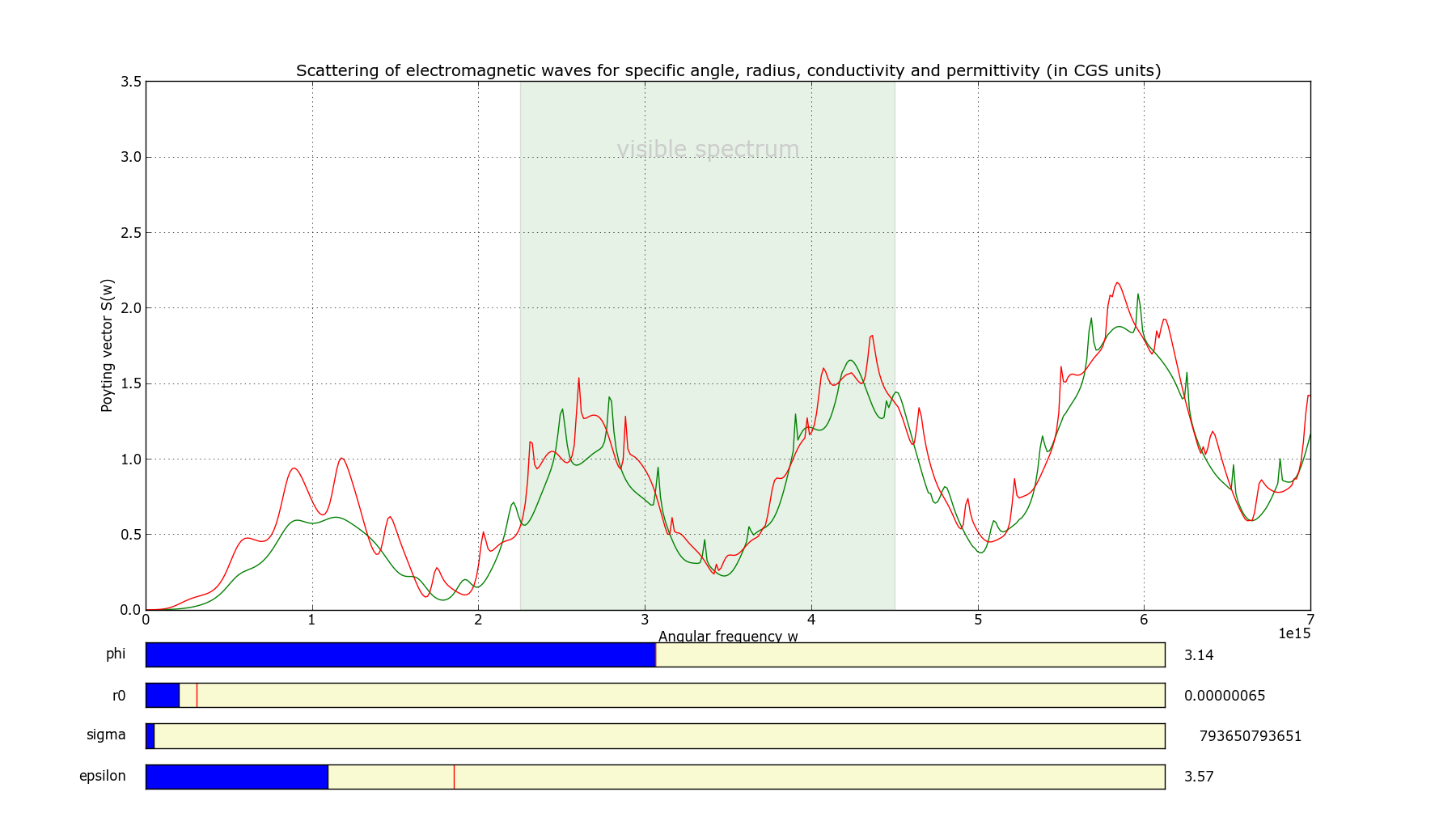}
	\caption{Dependence of scattering amplitude on frequency for two polarizations}
	\label{fig:EzEy}
\end{figure}
Authors  use formulas for scattered light integrated over the angle. But the observed scattered light only for $180^\circ$.
The dependence on frequency is rather strong for nanowires. This property was used in \cite{B} to propose a method which allows to determine precisely the radius of a wire. Our formalism gives a link to such physical constants as $\sigma,\epsilon$ determination by optical measurements.

\end{document}